\documentclass[12pt]{article}
\usepackage[margin=2cm]{geometry}
\usepackage{comment}
\usepackage{amsmath,amssymb,extarrows,graphicx,subfigure,setspace}
\usepackage{cite}
\usepackage{slashed}
\usepackage{color}
\usepackage[numbers, square, sort&compress]{natbib}
\makeatother

\numberwithin{equation}{section}

\newcommand{\be}{\begin{equation}}
\newcommand{\bea}{\begin{eqnarray}}
\newcommand{\eea}{\end{eqnarray}}
\newcommand{\ba}{\begin{array}}
\newcommand{\ea}{\end{array}}
\newcommand{\ee}{\end{equation}}

\begin{document}
\onehalfspacing
\noindent
\begin{titlepage}
\hfill
\vspace*{20mm}
\begin{center}
{\Large {\bf Aspects of Flat{/}CCFT Correspondence }\\
}

\vspace*{15mm} \vspace*{1mm} \textbf{{Reza Fareghbal$^{a,b}$,\;   Ali Naseh$^{b}$ }} 

 \vspace*{1cm}

$^a${\it  Department of Physics, 
Shahid Beheshti University, 
G.C., Evin, Tehran 19839, Iran.  \\ }
$^b${\it
School of Particles and Accelerators, 
 Institute for Research in Fundamental Sciences (IPM),  
 P.O. Box 19395-5531, Tehran, Iran }

\vspace*{.4cm}

{E-mails: {\tt r\_fareghbal@sbu.ac.ir, naseh@ipm.ir}}%

\vspace*{1cm}
\end{center}
\begin{abstract}
Flat/CCFT is a correspondence between gravity in asymptotically flat backgrounds and a field theory which is given by contraction of conformal field theory. In order to find a dictionary for Flat/CCFT correspondence  one can start from the AdS/CFT and take the contraction of CFT in the boundary as the dual description of   the flat-space limit (zero cosmological constant limit) of the asymptotically AdS spacetimes in the bulk side. In this paper we show  that the Cardy-like formula of CCFT$_2$ is given    by contraction of a proper formula in the  CFT$_2$. This formula is the modified Cardy formula which gives the entropy of inner horizon of BTZ black holes.

\end{abstract}
\end{titlepage}

\section{Introduction}

A possible way for generalizing the gauge/gravity duality for asymptotically flat spacetimes is using the AdS/CFT correspondence  and taking flat-space limit (zero cosmological constant limit) of asymptotically AdS spacetimes and trying to find a proper interpretation of this limit in the boundary CFT. A proposal for this method has been introduced in \cite{Bagchi:2010zz,Bagchi:2012cy} where the flat-space limit of the bulk corresponds to the contraction of boundary CFT. The contracted conformal field theory (CCFT) which is dual to asymptotically flat spacetimes has the same symmetry as the asymptotic symmetry of the bulk geometry.

The correspondence between the flat-space limit in the bulk and contraction of the dual boundary theory can be used in defining a dictionary for the Flat/CCFT correspondence. The first steps in this direction  have already done in papers \cite{Bagchi:2010zz,Bagchi:2012cy} which showed that the Bondi-Metzner-Sachs (BMS)  algebra \cite{BMS,bar-comp} as the asymptotic symmetry of asymptotically flat spacetimes at null infinity can be obtained by contraction of the Virasoro algebras as the symmetry of the two-dimensional CFT. Another important work in this context  is \cite{Bagchi:2012xr} where a Cardy-like formula for the CCFT$_2$ which is dual to the three-dimensional asymptotically flat spacetimes, has been introduced. The Cardy-like formula as an estimation of the CCFT  degeneracy of states, gives the exact entropy of the cosmological horizon of the three dimensional  cosmological solutions. These geometries are the shifted-boost orbifold of the three-dimensional Minkowski spacetimes and can be found by taking the flat-spce limit from the BTZ black holes.\cite{Cornalba:2002fi} .

Defining a field theory just by contracting a CFT has some advantages. Its correlation functions are given  by contraction of  correlation functions of the parent CFT \cite{GCA} . This fact can be also used for finding quasi local stress tensor of the asymptotically flat spacetimes. This idea has been followed in paper \cite{Fareghbal:2013ifa} for the three dimensional asymptotically flat spacetimes and a stress tensor was introduced which gives correct charges for the gravity solutions.  The stress tensor of \cite{Fareghbal:2013ifa} can be also used for computing  the BMS$_3$ charge algebra. 

The approach of using a  contracted CFT in  flat-space holography has been explored in many works. For example it can be used for finding higher-spin theories  in three-dimensional asymptotically flat spacetimes \cite{Afshar:2013vka, Gonzalez:2013oaa}. The reader can find a complete list of references in recent work \cite{Bagchi:2014ysa} where the current status of problem and also future directions have been mentioned. 

The point which we address  in this paper, is the  relation between the CCFT$_2$ Cardy-like formula  and the modified Cardy formula of CFT$_2$. The Cardy-like formula of CCFT has been derived in \cite{Bagchi:2012xr} by defining partition function of CCFT and demanding its invariance under some novel  modular transformations. In this paper we show that it  is simply  given by contracting a proper formula of the CFT. This formula is not the known Cardy formula which according to AdS$_3$/CFT$_2$ correspondence gives the correct entropy of the outer horizon of the BTZ black holes but it is the formula which gives the entropy of the inner horizon\cite{Detournay:2012ug, Castro:2012av}. We elaborate on this point in the main text by using  the flat-space limit  of the BTZ black holes in the bulk side. 

The organization of this paper is as follows: In the next section we briefly review the Flat/CCFT correspondence. The main point which we want to clarify in this section is the correspondence between the flat-space limit in the bulk and contraction of the CFT in the boundary.  In Section 3 we  find the Cardy-like formula of CCFT$_2$ by contracting its counterpart in the parent CFT. Last section is devoted to conclusions and possible future directions.      
     
\section{A Brief Review of the Flat{/}CCFT Correspondence  }
Let us consider three-dimensional Einstein gravity with negative cosmological constant,
\begin{equation}
S={1\over 16\pi G}\int\, d^3x \sqrt{-g}(R+{2\over\ell^2}).
\end{equation}
The flat-space limit for this theory is defined by taking zero cosmological constant limit which is given by  $\ell\to\infty$ limit. In order to make this limit well-defined we define the flat-space limit by using the  dimensionless parameter ${G/\ell}$ and sending it to zero while keeping $G$ fixed. At the level of metric, the flat-space limit is gauge dependent. An appropriate gauge that can capture properly the flat-space limit of the asymptotically locally AdS spacetimes  is known as  BMS gauge \cite{Isenberg:1983,Friedrich:1998wq,Barnich:2012aw}. 
   For the \textit{asymptotically locally} $AdS_{3}$ spacetimes
the general solution of  equations of motions  in the BMS gauge can be written as
\cite{Barnich:2012aw}
\bea\label{A.AdS.BMS}
ds^{2} = \left(-\frac{r^{2}}{l^{2}}+\mathcal{M}\right) du^{2} -2dudr+2\mathcal{N}dud\phi+r^{2}d\phi^{2},
\eea
where $u$ is the retarded time coordinate, and $\mathcal{M}$ and $\mathcal{N}$ are  functions of $u,\phi$ coordinates. Using the equations of motion, one then finds 
\bea
\partial_{u} \mathcal{M} =\frac{2}{l^{2}} \partial_{\phi}\mathcal{N},~~~~~~~2\partial_{u}\mathcal{N} =\partial_{\phi}\mathcal{M}.
\eea
 

It is shown in \cite{Fareghbal:2013ifa} that by proper expansion of functions $\mathcal{M}$ and $\mathcal{N}$ with respect to $G/l$, one can find the general asymptotically flat  metric as\footnote{We should emphasize that the flat-space limit in \cite{Fareghbal:2013ifa} is different from the modified Penrose limit defined in \cite{Barnich:2012aw} and the Grassmannian method introduced in \cite{123}. }
\bea\label{flat.BMS}
ds^{2} = M du^{2}-2dudr+2N dud\phi +r^{2}d\phi^{2},
\eea
where
\bea\label{MN}
M =\lim_{G/\ell\to 0}\mathcal{M}= \theta(\phi),~~~~~~~~~~~~~N =\lim_{G/\ell\to 0}\mathcal{N}= \chi(\phi)+\frac{u}{2}\theta^{\prime}(\phi).
\eea

The gauge (\ref{A.AdS.BMS}) implies an asymptotic symmetry algebra which is given by two copies of the Virasoro  algebra   \cite{Barnich:2012aw}:
\bea\label{CONFORMAL.algebra}
[\mathcal{L}_{m},\mathcal{L}_{n}] = (m-n) \mathcal{L}_{m+n},~~~~~[\bar{\mathcal{L}}_{m},\bar{\mathcal{L}}_{n}] = (m-n) \bar{\mathcal{L}}_{m+n},~~~~~[\mathcal{L}_{m},\bar{\mathcal{L}}_{n}] = 0.
\eea
In \cite{Barnich:2012aw}, the authors found the central extension of the surface charges algebra, computed with respect to AdS$_{3}$ background, with $c=\bar c=3 l/2 G$. 

Using \eqref{CONFORMAL.algebra} and explicit form of the generators $\mathcal{L}_{n}$ and $\bar{\mathcal{L}}_{n}$ introduced in  \cite{Barnich:2012aw}, one can easily check  that the new generators 
\bea\label{gravity.scaling}
L_{n} = \mathcal{L}_{n}-\bar{\mathcal{L}}_{-n},~~~~~~M_{n} = \frac{G}{l}(\mathcal{L}_{n}+\bar{\mathcal{L}}_{-n}),
\eea
in the  $G/l\rightarrow 0$ limit results in the
  BMS$_3$ algebra 
\bea\label{BMS.algebra}
[L_{m},L_{n}] = (m-n) L_{m+n},~~~~[L_{m},M_{n}] = (m-n) M_{m+n},~~~~
[M_{m},M_{n}] = 0,
\eea
which is the asymptotic symmetry of the three-dimensional asymptotically flat spacetimes\cite{bar-comp}. Moreover,  the  BMS$_3$ charge  algebra contains two central charges which are given by \cite{{bar-comp}}
\bea
C_{LL} = \lim_{\frac{G}{l}\rightarrow 0} \frac{c-\bar{c}}{12},~~~~~~~C_{LM} = \lim_{\frac{G}{l}\rightarrow 0} \frac{G}{l} \left(\frac{c+\bar{c}}{12}\right). 
\eea

 Now let us focus on the boundary side and look for the equivalent procedure for  the flat-space limit in the dual boundary  CFT$_2$ of the asymptotically AdS$_3$ spacetimes.

In order to answer this question let us have a closer look at the generic solution \eqref{A.AdS.BMS} and try to find the conformal boundary for an arbitrary large $\ell$. The metric of the conformal boundary is the same for all $\mathcal{M}$ and $\mathcal{N}$ and is given by:
\begin{equation}
ds^2={r^2\over G^2}\left(-{G^2\over\ell^2}du^2+G^2 d\phi^2\right).
\end{equation}
Thus, $\ell$ can be absorbed in the definition of new time $t={G\over\ell}u$. The dual CFT lives on a cylinder with coordinates $\{t,\phi\}$ and radius $G$. It is clear that taking the flat-space limit ,${G\over\ell}\to 0$, is equivalent to contract time as $t\to \epsilon t$ with $\epsilon \to 0$, thus one may guess  that the dual of asymptotically flat spacetimes is a CCFT. In two dimensions the symmetry of CCFT is isomorphic to the Galilean conformal algebra(GCA)\cite{GCA}. In \cite{Bagchi:2010zz,Bagchi:2012cy}, it has been argued that one can obtain the full GCA in two dimensions by contracting the symmetries of the two-dimensional CFT.  In this approach the generators of the GCA and the parent CFT  and also their central charges  are related by
\begin{eqnarray}\label{GCAgC}
M_n&=&\lim_{\epsilon\to0}\epsilon\left(\mathcal{L}_n+\mathcal{\bar L}_{-n}\right),\qquad L_n=\lim_{\epsilon\to0}\left(\mathcal{L}_n-\mathcal{\bar L}_{-n}\right),\\
C_{LL} &=& \lim_{\epsilon\to 0} \frac{c-\bar{c}}{12},~~~~~~~C_{LM} = \lim_{\epsilon \to 0} \epsilon (\frac{c+\bar{c}}{12}).
\end{eqnarray}

  Therefore,  the connection between the flat-space limit in the bulk side and contraction of CFT in the boundary side is  correct at the level of symmetries and one may propose a  dual field theory for the asymptotically flat spacetimes which is a CCFT. We call this duality Flat/CCFT\footnote{Originally this correspondence was coined as BMS/GCA \cite{Bagchi:2010zz,Bagchi:2012cy} but since for  bulk dimensions  greater than three the BMS algebra is not exactly GCA while it is still related to a contracted conformal algebra, we  call it  Flat/CCFT correspondence. }\footnote{The coordinate which must be contracted can be determined by arguments related to those used for the conformal boundary which has been done in this section. It was time coordinate which needed to be contracted for the dual of AdS written in the BMS gauge. As discussed in  \cite{Fareghbal:2014oba}, for Rindler-AdS one should contract x-coordinate in order to find dual of Rindler spacetime.}.

The Falt/CCFT correspondence can be used for finding the quasi local   stress tensor of asymptotically flat spacetimes. We expect the same dictionary as the  AdS/CFT correspondence i.e. one-point function of  energy-momentum operator of the dual CCFT is the stress tensor of the bulk theory. This connection has been addressed in \cite{Fareghbal:2013ifa} where we have introduced the stress tensor of the asymptotically flat spacetimes  using the Flat/CCFT correspondence. The key-point which has been used in \cite{Fareghbal:2013ifa} is that, therein, the authors used the previously known results about the one-point functions of the GCA and then constructed the flat-space stress tensor\footnote{The holographic renormalization  for the asymptotically flat spacetimes has been also worked out in papers \cite{Costa:2013vza} and \cite{Detournay:2014fva} but non of them explored the connection between flat-space holography and CCFTs. }. The energy-momentum of field theories which  arise by contraction of conformal field theories is given by using energy-momentum tensor of original CFT. For example, for field theories with Galilean conformal symmetry, this connection has been worked out in papers \cite{GCA}.  The CCFT which is dual description of asymptotically flat spacetimes written in the BMS gauge, is defined by contracting time in the original CFT. Thus One would expect that the CCFT Hamiltonian is related to the Hamiltonian of the parent CFT by contracting time, while the momentum operator of both theories are the same, since the $x$-coordinate does not affect by the limiting procedure.

 The flat-space stress tensor has been worked out in \cite{Fareghbal:2013ifa}\footnote{The results of \cite{Fareghbal:2013ifa} predict a symmetric structure for the CCFT energy-momentum (EM) tensor. On the other hand one may expect a non-symmetric  EM tensor due to  absence of  Lorentz symmetry for the CCFT. We should emphasize that the symmetric structure of CCFT energy-momentum tensor in the current case is a direct consequence of zero off-diagonal terms for the components of the parent CFT energy-momentum  tensor   written in the light-cone gauge. Any improvement to   the generic case requires a clear understanding of the EM tensor of a theory with Galilean conformal symmetry which may be achieved  along the lines of   recent papers \cite{Christensen:2013rfa}-\cite{Hartong:2014pma}. }. 
\bea\label{flat.stress.tensor}
\tilde{T}_{uu} =\frac{M}{16\pi G^{2}},~~~~~~~~~~\tilde{T}_{u\phi} = \frac{N}{8\pi G^{2}},~~~~~~~~~\tilde{T}_{\phi\phi} = \frac{M}{16\pi},
\eea
where the functions $M$ and $N$ are given by (\ref{MN}). 

In  \cite{Fareghbal:2013ifa} the above   stress tensor \eqref{flat.stress.tensor} is used  to find 
 conserved charges of symmetry generators $\xi^{\mu}$ by using the Brown and York definition \cite{Brown:1992br}
\bea
{Q}_{\xi} = \int_{\Sigma} d\phi \sqrt{\sigma} \upsilon^{\mu}\xi^{\nu}\tilde{T}_{\mu\nu},
\eea
where $\upsilon^{\mu}$ is the unit timelike vector normal to $\Sigma$.  
The variation of charges under the symmetry generators yields the BMS$_3$ algebra with exactly the same central extension as \cite{bar-comp}. Using the results of \cite{Fareghbal:2013ifa}, we can find the corresponding charges of the generators $M_0$ and $L_0$ as 
\bea\label{vacuum.charges}
Q_{M_{0}} = -\frac{1}{8G},~~~~~~~~Q_{L_{0}} = 0. 
\eea
We will use these  charges in the next section to compute the entropy of the cosmological solution using the CCFT Cardy-like formula.

\section{A Cardy-Like Formula for the CCFT by Contraction of the Modified Cardy Formula }
Interpreting  black holes entropy as  degeneracy of  microstates of  dual theory is an important issue which must be addressed carefully in any gauge/gravity correspondence. The Flat/CCFT correspondence as a duality between gravity in the asymptotically flat spacetimes and Contracted CFTs must provide a clear understanding for the entropy of the asymptotically flat black holes  using microstates of the dual CCFT. 

Let us concentrate on the Flat$_3$/CCFT$_2$  and consider the three-dimensional asymptotically flat spacetimes. It was shown in \cite{Ida:2000jh} that no  asymptotically flat black hole exists in the three-dimensional Einstein Gravity.  However, there are other interesting solutions in three dimensional Einstein gravity which have cosmological horizon. They are called  cosmological solutions and are given by the following metric:
\begin{equation}\label{eq:shiftorb}
ds^2=\hat r_+^2 d t^2-\frac{r^2\,dr^2}{\hat r_+^2 (r^2-r_0^2)}+r^2 d\phi^2-2 \hat r_+ r_0
 dt d \phi, \;\;\quad
\end{equation}
These solutions are characterized   by two parameters $\hat r_+$ and $r_0$ which are related to the mass $M$ and the angular momentum $J$ as $   \hat{r}_+=\sqrt{8GM}  $ and $r_0= \sqrt{\frac{2G}{M}} |J|$. Moreover, $r=r_0$ is the radius of  the cosmological horizon and one  can define  the entropy of the cosmological solution by using the area of the cosmological horizon as 
\begin{equation}\label{entropy of CS}
 S={A\over 4 G}={\pi r_0\over 2 G}.
 \end{equation} 
It was shown that the solution \eqref{eq:shiftorb} is the boost-shift orbifold of the three-dimensional Minkowski spacetimes and it can be obtained by taking the flat-space limit of the  BTZ black hole\cite{Cornalba:2002fi}. The connection between BTZ black holes and cosmological solutions is important for us in this paper. The idea that the  flat-space limit in the bulk corresponds to the contraction of CFT in the boundary can be used to  find a Cardy-like formula for the states of CCFT which yields the entropy of the cosmological solution. In fact this idea was  first used in paper \cite{Bagchi:2012xr} where a Cardy-like formula was introduced for the CCFT$_2$. The idea of contraction of time for the CFT has been entered in an unusual  transformation of the modular parameters of the CCFT. Invariance of the partition function under these new modular transformations along with a  saddle point  approximation resulted in a  formula for the degeneracy of states which correctly reproduces entropy of the cosmological horizon\cite{Bagchi:2012xr}.  In this paper we want to argue that the Cardy-like formula of CCFT$_2$ can be obtained by taking a direct contraction of the modified Cardy formula of CFT. Before making this connection we need to have a closer look at the bulk and study the  flat-space limit of the BTZ black hole.

The cosmological solution  is given by taking the flat-space limit of the BTZ black hole with metric 
\be\label{BH metric}
 ds^2=-{(r^2-r_+^2)(r^2-r_-^2)\over r^2\ell^2}dt^2+{ r^2\ell^2\over(r^2-r_+^2)(r^2-r_-^2)}dr^2+r^2\left(d\phi+{r_+r_-\over \ell r^2}dt\right)^2,
 \ee
 where
 \be\label{horizon radia}
 r_\pm=\sqrt{2G\ell(\ell M+J)}\pm\sqrt{2G\ell(\ell M-J)}
 \ee
are the radii of the horizons and  $M$ and $J$ are related to the mass and the angular momentum of the
black hole. It is easy to see that $r_0$ in the cosmological solution is given by taking $\ell\to\infty$ limit of $r_-$ but $r_+$ goes to infinity in the flat-space limit. On the other hand according to AdS$_3$/CFT$_2$ correspondence, the Cardy formula of CFT,
\bea\label{original.cardy.formula}
S = 2\pi\sqrt{\frac{c}{6}\mathcal{L}_{0}}+2\pi\sqrt{\frac{\bar{c}}{6}\bar{\mathcal{L}}_{0}}.
\eea    
results in the correct entropy of the outer horizon $r_+$. We can rewrite  \eqref{original.cardy.formula}  using \eqref{GCAgC} and take $\epsilon\to 0$ limit which corresponds to $\ell\to\infty$ limit in the bulk. The final answer diverges as expected because in the bulk side the entropy of the outer horizon diverges in the $\ell\to \infty$ limit. If we want to find a Cardy-like formula which gives the entropy of the cosmological horizon we should contract a  formula in the CFT  which corresponds to the  entropy of the inner horizon of the BTZ. 

The entropy of the inner horizon of the BTZ black hole in terms of the dual CFT parameters, can be written as \cite{Detournay:2012ug, Castro:2012av}:
\bea\label{inner original.cardy.formula}
S = \left|2\pi\sqrt{\frac{c}{6}\mathcal{L}_{0}}-2\pi\sqrt{\frac{\bar{c}}{6}\bar{\mathcal{L}}_{0}}\right|.
\eea
Now if we use \eqref{GCAgC} and take the $\epsilon\to 0$ limit the final answer is well-defined and is exactly the Cardy-like formula which was introduced in \cite{Bagchi:2012xr} i.e. 
\begin{equation}\label{Cardy-like}
  S =    2\pi\bigg( C_{LL}
\sqrt{\frac{M_0}{2C_{LM}}} + L_0 \sqrt{\frac{C_{LM}}{2M_0}}
\bigg).
  \end{equation} 
This is another check that flat limit in the bulk corresponds  to contraction of CFT in the boundary.  
 
 The Cardy-like formula for the CCFTs can also be given in another form. The point is that there is another form for the Cardy formula (see appendix of \cite{Detournay:2012pc} and the references therein). Now one can use this new formulation and take limit from it.
 
  The alternative form of the Cardy formula which gives  the degeneracy of states of the CFT is given by 
\bea\label{another Cardy}
S_{CFT} = 2\pi \sqrt{-({H}_{vac}+{J}_{vac})({H}+{J})} +2\pi\sqrt{-({H}_{vac}-{J}_{vac})({H}-{J})}.
\eea 
 States are  characterized by the eigenvalues of the Hamiltonian and the momentum  $( H,J)$ and "vac"  denotes the vacuum state. Using the AdS/CFT correspondence, the above formula \eqref{another Cardy} gives the entropy of the outer horizon of the BTZ black holes \cite{Detournay:2012pc}.
However, if we want to find a similar formula for the CCFT, similar to the previous subsection, we should look for a formula in terms of  CFT parameters which gives the correct entropy of inner horizon of BTZ black holes. Using  \cite{Detournay:2012pc}, one can easily check that the  formula 
\bea\label{another Cardy for inner}
S = 2\pi \sqrt{-({H}_{vac}+{J}_{vac})({H}+{J})} -2\pi\sqrt{-({H}_{vac}-{J}_{vac})({H}-{J})},
\eea
results in an entropy which matches with the entropy of inner horizon.

Now we can  contract \eqref{another Cardy for inner}. As discussed in section 2, the Hamiltonian of CCFT is given by contracting the corresponding Hamiltonian of CFT but momenta  are the same for both of the theories. If we use this correspondence between $( H,  J)$  of CFT and $(\tilde H,\tilde J)$ of CCFT, we can write
\bea\label{alter..cardy.CCFT}
S_{CCFT} &=& \lim_{\epsilon\rightarrow 0} \left[2\pi \sqrt{-(\frac{\tilde H_{vac}}{\epsilon}+\tilde J_{vac})(\frac{\tilde H}{\epsilon}+\tilde J)} -2\pi\sqrt{-(\frac{\tilde H_{vac}}{\epsilon}-\tilde J_{vac})(\frac{\tilde H}{\epsilon}-\tilde J)}~\right],\cr\nonumber\\
&=& \lim_{\epsilon\rightarrow 0} \left[2\pi~\frac{-\frac{2}{\epsilon} (\tilde H_{vac}\tilde J+\tilde J_{vac}\tilde H)}{\frac{2}{\epsilon}\sqrt{-\tilde H_{vac}\tilde H}}\right] = 2\pi \sqrt{-\frac{\tilde H_{vac}}{\tilde H}}\tilde J +2\pi \sqrt{-\frac{\tilde H}{\tilde H_{vac}}}\tilde J_{vac},
\eea
 Using (\ref{vacuum.charges}), in the bulk side we have
\bea
Q_{M_{0}} = \tilde H_{vac}= -\frac{1}{8G},~~~~~~Q_{L_{0}} = \tilde J_{vac} = 0,
\eea
where  $\tilde H_{vac}$  exactly matches with the  free-energy of the cosmological solution that can be obtained from the Euclidean on-shell action \cite{Bagchi:2013lma}. Therefore, from \eqref{alter..cardy.CCFT} the entropy of cosmological solution  is found to be
\bea
S = \frac{\pi}{2G} J \sqrt{\frac{2G}{M}}.
\eea
Using $r_{0} = J \sqrt{\frac{2G}{M}}$, we end up with the known result of the entropy for the cosmological solution\cite{Bagchi:2012xr},
\bea
S = \frac{\pi r_{0}}{2G}.
\eea

\section{Conclusion}

This work is another check for the correspondence between the flat-space limit in the bulk  and contraction  of the boundary field theory. We considered the three-dimensional Einstein gravity which admits asymptotically flat geometries, namely cosmological solutions. The asymptotic symmetry group of the asymptotically flat solutions at null infinity is infinite dimensional. The next step is studying the four-dimensional asymptotically flat spacetimes which also have infinite dimensional asymptotic symmetry group\cite{bar-comp}. The lessons from the three-dimensional analysis can be used for finding the quasi local stress tensor of the four-dimensional spacetimes and one can also calculate the BMS$_4$ charge algebra  using our approach and get insight about the possible central extension of the BMS$_4$ algebra. 

Moreover, a  Cardy-like formula in the form \eqref{alter..cardy.CCFT}  may exist for the CCFT$_3$ which also has  infinite dimensional symmetry. We should note that the formulas  \eqref{inner original.cardy.formula} and \eqref{another Cardy for inner} which gives the  inner horizon entropy are phenomenological observation and there is not a proof for them in the CFT side. We just wrote the  inner horizon entropy  of the BTZ in terms of the CFT parameters but what  this counting means, is an open question.

\section*{Acknowledgement}
The authors  would like to especially  thank Mahmoud Safari and Seyed Morteza Hosseini  for their comments on the manuscript and Stephane Detournay for useful comments. We also thank Arjun Bagchi,  Daniel Grumiller and Joan Simon for a continuing  collaboration  in the flat-spacetime Holography project.\\

\textbf{\textit{Note added:}} When this paper was ready for submission, we became aware that Max Riegler also  has pointed out  the limit of the modified  Cardy  formula which results in   \eqref{Cardy-like}\cite{Riegler:2014bia}.

\end{document}